\documentclass[preprint]{aastex}

\usepackage{epigraph}

\usepackage{graphicx}
\usepackage{abbreviations}
\usepackage{natbib}
\usepackage{amssymb}

\newcommand{\be}{\begin{equation}}
\newcommand{\ee}{\end{equation}}
\newcommand{\nn}{\mbox{} \nonumber \\ \mbox{} }
\newcommand{\ba}{\begin{eqnarray}}
\newcommand{\ea}{\end{eqnarray}}

\newcommand{\Alfven}{ Alfv\'{e}n }

\newcommand\eg{\textit{e.g.,\ }}

\newcommand{\Bf}{{magnetic field}}
\newcommand{\Bfs}{{magnetic fields}}
\newcommand{\Ef}{{electric  field}}
\newcommand{\Efs}{{electric fields}}

\begin{document}

\title
{What we recently  learnt about Crab: structure of the wind, the shock, flares and reconnection}
\author
{
Maxim Lyutikov \\
Department of Physics and Astronomy, Purdue University, West Lafayette, IN 47907-2036, USA
}


\begin{abstract}
We can probe observationally and reproduce theoretically the most detailed properties of the Crab Nebula nearest to the pulsar - The Inner Knot.    The tiny knot is indeed a bright spot on the surface of a quasi-stationary magnetic relativistic shock that efficiently accelerates particles. It required that the part of the wind that produces the Inner Knot has low magnetization; thus, it is not a site of gamma-ray flares.

We develop a model of  particle acceleration during explosive reconnection events in relativistic highly magnetized plasma and apply the model to explain the Crab gamma-ray flares. Particles are efficiently accelerated by charge-starved DC-type electric fields during initial stages of magnetic flux merges.  Thus,  magnetic reconnection is an important, and possibly dominant process of particle acceleration in high energy astrophysical sources.
\end{abstract}


\maketitle

\epigraph{Astronomy consists of two parts: 
    the Crab Nebula and all the rest.}{Attributed to I.S. Shklovsky}

\section{Introduction}

Understanding the nature of the Crab Nebula  was one of the major achievements of Iosif Shklovsky  \cite[Shklovskii, I.S., ÒOn the Nature of the Crab NebulaÕs Optical Emission,Ó Doklady Akad. Nauk SSSR 90, 983 (1953),  in Russian; translated with commentary in ][]{1979sbaa.book.....L}. One might argue that the  idea that the continuum emission is generated by a population of relativistic electrons via synchrotron emission gave birth to a  new field  -  relativistic astrophysics.  Only few years before Shklovsky' paper  Minkowski wrote  "The only physically justified assumption is that the continuous spectrum is produced by free-free and free-bound transitions" \citep{1942ApJ....96..199M}.

The Crab pulsar and its pulsar wind nebula (PWN) remain the prime targets for 
high energy astrophysical research. In many ways,  the Crab Nebula  is the paragon of astrophysical high energy source - many of the current models
of Active Galactic Nuclei and Gamma Ray Bursts are based on what we have 
learned from the studies of the Crab.  The  recent detection of flares
from the Crab Nebula by AGILE and Fermi satellites
\citep{2011Sci...331..736T,2011Sci...331..739A} have brought this object 
into the ``focal point'' once again.  Their extreme properties seem impossible
to explain within the standard theories of non-thermal particle acceleration 
and require their overhaul with important implications to 
high energy astrophysics in general 
\citep[e.g.][]{2010MNRAS.405.1809L,2012MNRAS.426.1374C,2012ApJ...746..148C,2014RPPh...77f6901B,2016arXiv160305731L}

\section{Crab Inner knot}

In the MHD models of the Crab Nebula, the super-fast-magnetosonic
relativistic wind of the Crab pulsar terminates at a reverse shock
\citep{reesgunn,kc84}.  
However, finding the shock in the images of the Crab Nebula has 
not been a straight-forward matter - there seem to be no sharp feature which can 
be undoubtedly identified with the shock surface. In their seminal paper,  
\citet{kc84} discuss the under-luminous region hosting the Crab pulsar and surrounded 
by the optical wisps as an indicator of the shock presence. After the discovery of the 
inner X-ray ring by Chandra 
\citep{2000ApJ...536L..81W,2002ApJ...577L..49H}, the ring is often referred to as 
the termination shock and yet 
this feature looks much more like a collection of knots than a smooth surface.  

The wind from an oblique rotator should have the so-called 
striped zone where the orientation of magnetic field alternated on the scale of 
the pulsar period.  The magnetic energy associated with these stripes can 
be dissipated at the termination shock and converted into the energy of the wind 
particles \citep{2003MNRAS.345..153L,2011ApJ...741...39S}.

\begin{figure}[h!]
\centering
\includegraphics[width=0.64\textwidth]{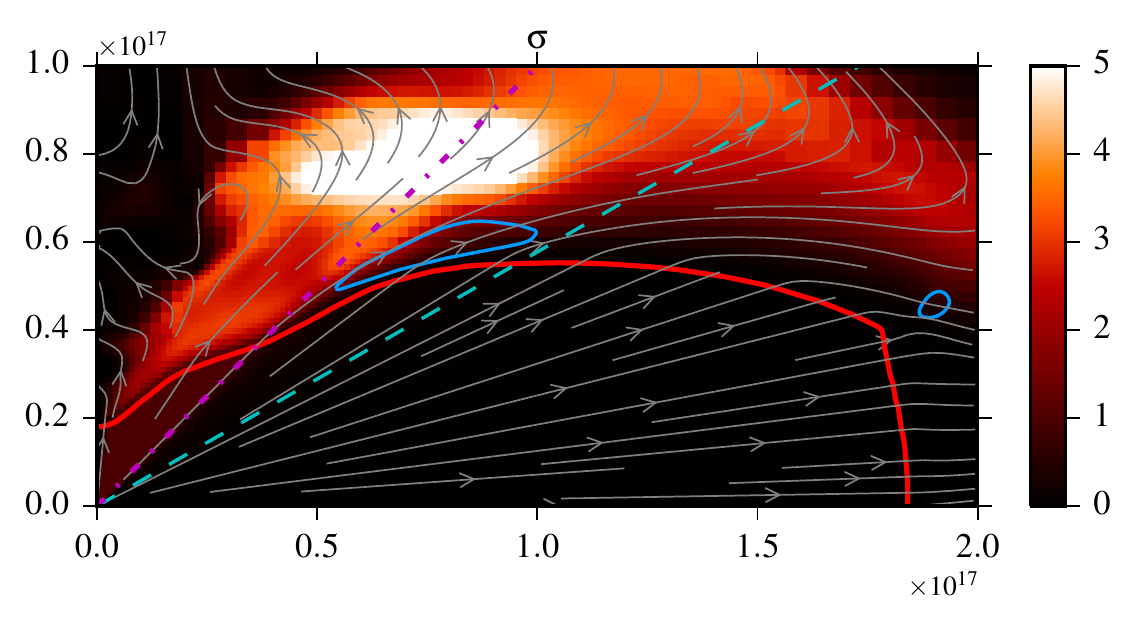}
\includegraphics[width=0.35\linewidth]{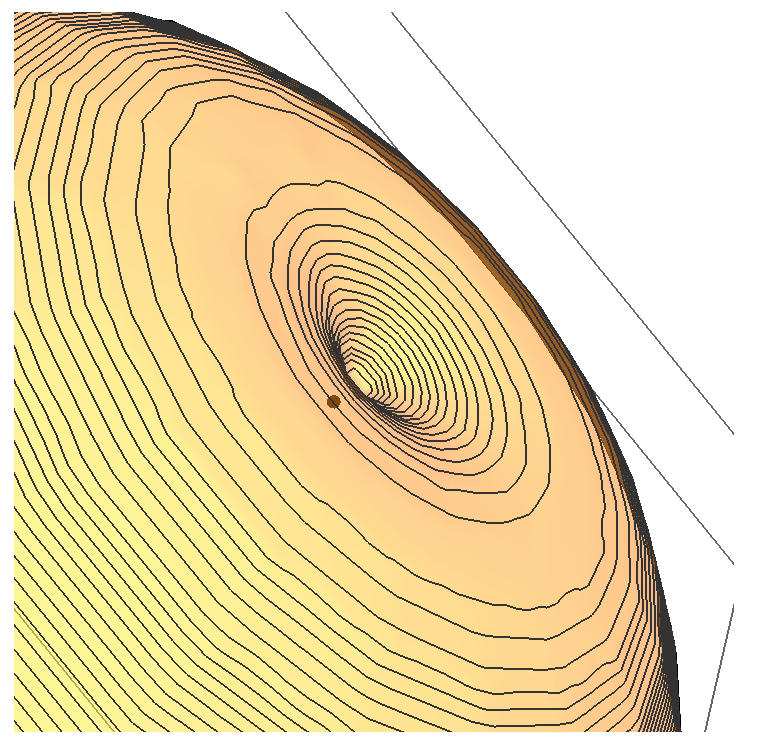}
\caption{Left Panel: Zoom-in on the central part of global PWN simulation  \citep{2016MNRAS.456..286L}. The dashed line
is the line of view and the blue curves show the regions of enhanced
observed emissivity.The dot-dashed line separates the high and low
magnetization regions of the wind.  The arrowed lines are the instantaneous stream lines.  Red line if the position of the shock. There is a clearly visible region of high magnetization at intermediate latitudes.  Right Panel:
View of the polar region of the termination shock for $f(\theta)=\sin^2\theta$ and 
the viewing angle $\theta_{ob}=60^\circ$; the pulsar position is shown by the dot.}
\label{point}
\end{figure}

Given the highly anisotropic nature of the
wind, the termination shock is squashed along the polar direction, Fig. \ref{point} and can be
highly oblique with respect to the upstream flow
\citep{2002MNRAS.329L..34L}. Downstream of the shock, the flow can
still be relativistic and its emission subject to strong
Doppler beaming.

The computer simulations of the Crab nebula and its
radiation \citep{KomissarovLyubarsky} (see also more advanced more advanced 2D \citep{2009MNRAS.400.1241C} and 3D
\citep{2014MNRAS.438..278P} simulations)  revealed the presence of a very
bright compact feature in the synthetic synchrotron maps, highly
reminiscent of the HST knot 1 of the Crab Nebula located very close
to the pulsar \citep[also called the inner knot, ][]{1995ApJ...448..240H}.  
(In these simulations, the termination shock was treated as source of 
synchrotron electrons with power-law energy spectrum, which then were carried out 
into the nebula by the shocked wind plasma.)

Recently, a targeted multi-wavelength study of the Crab's inner knot
has been conducted by \cite{2015ApJ...811...24R} in order to check if it shows any 
activity correlated with the gamma-ray flares. 
Although no such correlation has been found, the optical 
data reveal the structure and temporal evolution of the knot with unprecedented 
detail. 

 \cite{2016MNRAS.456..286L}  investigated if the data are consistent with the MHD-shock model of 
the knot using simple analytical and semi-analytical tools.  In particular, they combined 
the theoretical shape of the shock with the oblique shock jumps in order to obtain 
the Doppler-beaming of the post shock emission and use this to determine the location, 
the shape and the brightness distribution of the knot. 
The model,  Fig. \ref{mapSigma}, successfully explains a number of its observed properties: 

\begin{figure}[h!]
\includegraphics[width=0.45\linewidth]{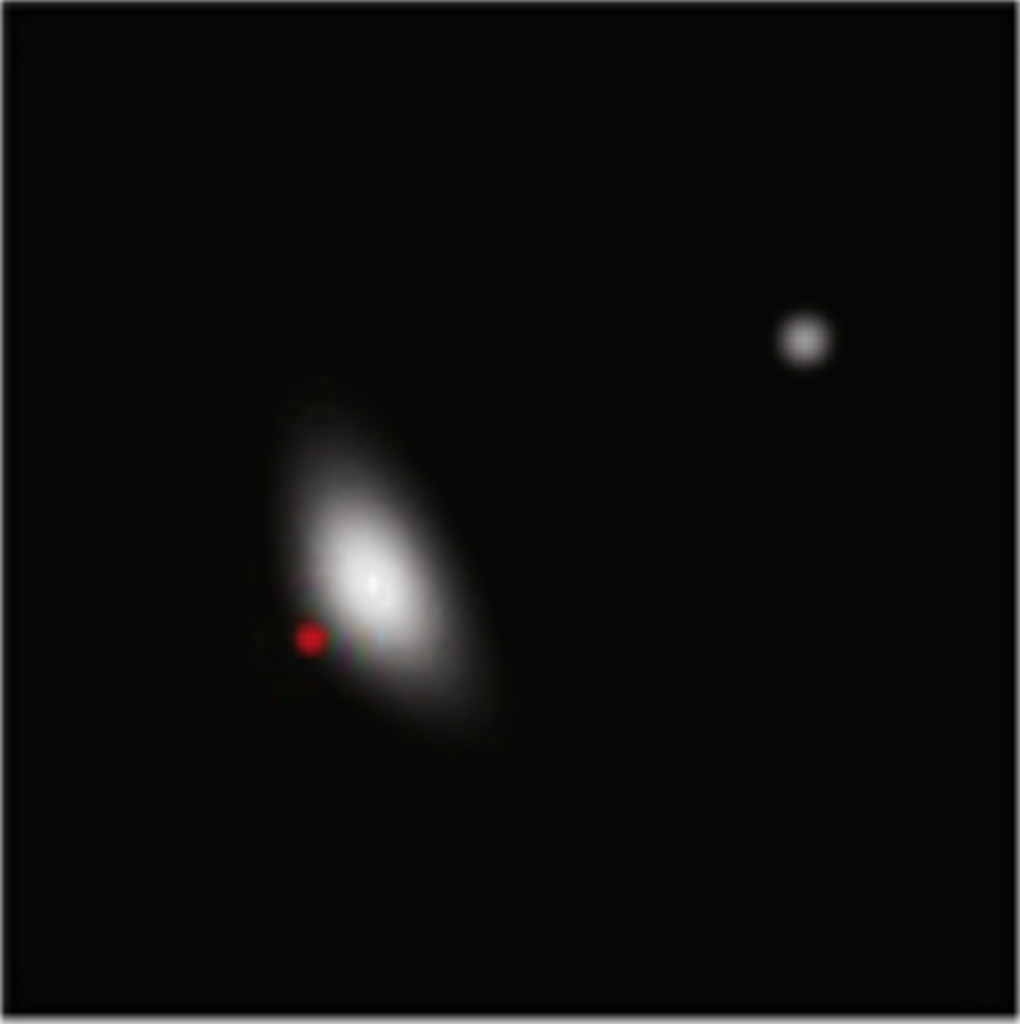}
\includegraphics[width=0.55\linewidth]{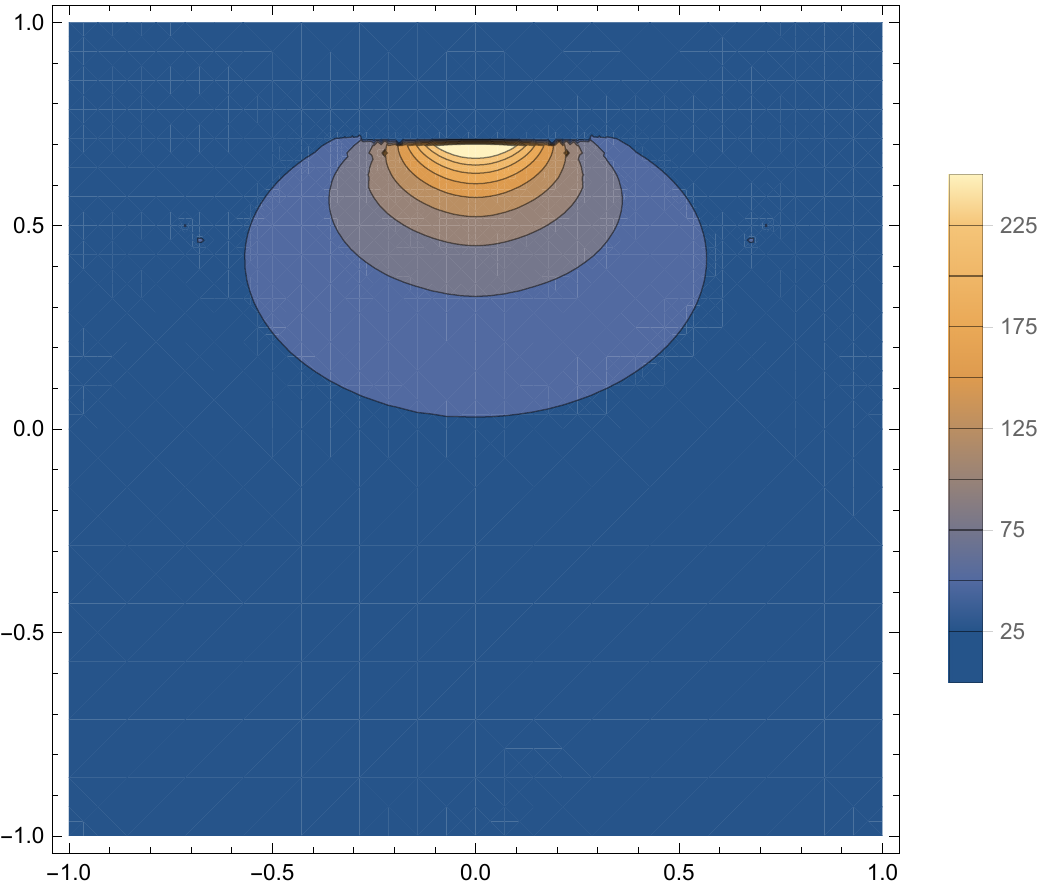}
\caption{Left Panel: one of the images from  \citet{2015ApJ...811...24R}. Right panel: theoretical  emission map for  magnetization parameters, $\sigma
  =0$ \cite{2016MNRAS.456..286L}. The pulsar is located at the
  origin.}
\label{mapSigma}
\end{figure}

\begin{description} 

\item{\it Location:} The knot is located on the same side of the pulsar as the Crab jet,   
along the symmetry axis of the inner nebula, and on the opposite side as the brighter 
section of the Crab torus. This is a direct consequence 
of the termination shock geometry and the Doppler-boosting.  

\item{\it Size:} The knot size is comparable to its separation from the pulsar. 
This also follows from the shock geometry and the Doppler-beaming. The anisotropy 
of the proper synchrotron emissivity, which vanishes along the magnetic field 
direction in combination with the relativistic aberration of light is another 
significant factor. Only models with low magnetization of the post-shock flow,
with the effective magnetization parameter of the wind $\sigma_1 < 1$ agree with 
the observations. 

\item{\it Elongation:} The knot is elongated in the direction perpendicular to the symmetry 
axis. This is because the knot emission comes from the region where the shock surface is 
almost parallel to the line of sight.

\item{\it Polarization:} The knot polarization degree is high, and 
the electric vector is aligned with the symmetry axis. This come due to the fact that 
the post-shock magnetic field is highly ordered in the vicinity of the termination shock 
and azimuthal. In the model, the relativistic aberration of light leads to a noticeable 
rotation of the polarization vector along the knot and this prediction could be tested in 
future polarization observations. Accordingly, the polarization degree of the integral  
knot emission depends on the integration area - the bigger the area the smaller the 
degree is.        

\item{\it Luminosity:} Taking into account Doppler beaming, the observed radiative 
efficiency of the inner knot is consistent with efficient particle acceleration at the 
termination shock and the knot's magnetic field of one milli-Gauss strength, 
which is a reasonable value for the inner Crab Nebula. 

\item{\it Variability:} The knot flux is anti-correlated with its separation from the pulsar. 
In the numerical simulations, the termination shock is found to be highly unsteady, changing its
size and shape. As the shock moves away from the pulsar, so does the knot region, which leads 
to lower magnetic field and hence lower emissivity. Another outcome of the shock variability
in the MHD simulations is the emission of wisps and hence one expects both the processes 
to occur on the same time-scale, which is consistent with the observations.        
  
  \item{\it Relation to Crab $\gamma$-ray flares:} The model requires that the sector of the wind that produces the inner knot has low magnetization $\sigma \ll 1$. Thus, it is not a region where flares originate since flares require highly magnetized medium, \S \ref{flares}.
\end{description}

Our results may have a number of important implications to the astrophysics of 
relativistic plasma in general and that of PWN in particular. They show that 
the termination shock of the relativistic wind from the Crab pulsar is a 
reality and that this shock is a location of efficient particle acceleration.      
The strong Doppler-beaming of the emission from the shock explains why this 
shock has been so elusive. Only the emission from a small patch on the shock 
surface, the inner knot, is strongly Doppler-boosted and hence prominent. 
For most of the shock, its emission is beamed away from the Earth and hence 
difficult to observe.

The shock model of the inner knot allows us to constrain the parameters of 
the wind from the Crab pulsar. Taken directly, the model requires the wind to 
be particle-dominated, $\sigma_1 < 1$ , at least at the polar latitudes of 
$40^\circ-60^\circ$. However, in the case of a striped wind, its termination 
shock can mimic that of a low $\sigma$ flow even when the actual wind magnetization 
is extremely high \citep{2003MNRAS.345..153L}. In this context, the   
magnetic inclination angle of the Crab pulsar should be above $45^\circ$, 
which means that  most of the Poynting flux of the Crab wind is converted into 
particles, if not in the wind itself then at its termination shock 
\citep{2013MNRAS.428.2459K}.  This is in agreement with the results of numerical 
simulations, which can reproduce the observed properties of the inner Crab 
Nebula extremely well in models with moderate wind magnetization 
\citep{2014MNRAS.438..278P}. However, the polar region of a pulsar 
wind is free of stripes and can still inject highly magnetized plasma into 
its PWN.  
 
 \section{Crab gamma-ray flares}
 \label{flares}
 
 The detection of  flares from  Crab Nebula by AGILE and Fermi satellites  \citep{2011Sci...331..736T,2011Sci...331..739A,2012ApJ...749...26B} is one of the most astounding discoveries in high energy astrophysics. 
The unusually short durations, high luminosities, and high photon energies of the Crab Nebula gamma-ray flares require reconsideration of our basic assumptions about the physical processes responsible for acceleration of the highest-energy emitting particles in the Crab Nebula, and, possibly in other high-energy  astrophysical sources. 

The Crab flares are characterized by an increase of  gamma-ray flux above 100 MeV by a factor of few or more on 
the day time-scale.  This energy corresponds to the high end of the Crab's synchrotron spectrum.   
Most interestingly,  in the other energy bands nothing unusual has been observed  during the flares so far \citep{2013ApJ...765...56W}. This suggests that the physical processes behind the flares lead to a dramatic increase of the highest energy population of relativistic electrons in the nebula, whereas lower energy population remains largely unaffected.  The short duration of flares indicate explosive and highly localised events. 
   
Most importantly, the peak of the flare spectrum approaches and even exceeds the maximal rest-frame synchrotron photon energy  \citep{1996ApJ...457..253D,2010MNRAS.405.1809L,2012MNRAS.426.1374C}. 
Balancing the synchrotron energy losses in the magnetic field $B$ against     
the energy gain via acceleration in the electric field of strength $E=\eta B$ leads to the upper limit 
of the synchrotron photon energy 
$$
\epsilon_{\rm max} \sim \eta   \hbar  { m c^3 \over e^2} \approx 100 \mbox{ MeV}
\label{emax}
$$
The high conductivity of astrophysical plasma ensures that for typical accelerating electric field  $\eta < 1$.  The fact that the flare spectrum extends beyond this limit pushes $\eta$ towards unity, which implies energy gain on the scale of the gyration period.   This  practically {\it excludes stochastic acceleration mechanisms} in general and the shock acceleration in particular. In principle, strong Doppler boosting could somewhat reduce this constraint but the  lack of observational evidence for ultra-relativistic macroscopic motion inside the nebula  makes this unlikely.

 A widely discussed alternative to the shock acceleration mechanism is the particle acceleration accompanying  
 magnetic reconnection.  It is well known that magnetic reconnection can lead to explosive release of magnetic energy, e.g. in solar flares.  However, properties of plasma in the Crab Nebula, as well as magnetospheres of pulsars and magnetars, pulsar winds, AGN and GRB jets and other targets of relativistic astrophysics,  are very different from those of more conventional Solar and laboratory plasmas \citep{2013SSRv..178..459L}.  In particular, the energy density of magnetic field can exceed not only the thermal energy density but also the rest mass-energy density of plasma particles.  In order to quantify such a strong magnetization, it is convenient to use the relativistic magnetization parameter 
\be
\sigma =  \frac{B^2}{4 \pi w}
\ee
where $w=\rho c^2 + (\hat{\gamma}/\hat{\gamma}-1)p$ is the relativistic enthalpy, which includes the rest mass-energy density of plasma. In traditional plasmas this parameter is very small but in relativistic astrophysics
$\sigma \gg 1$ is quite common.  This parameter is uniquely related to the \Alfven speed $v_A$  via 
$$ 
(v_A/c)^2 = \sigma /(1+\sigma) \,.
$$

We developed a model of   particle acceleration   in  explosive reconnection events in relativistic
magnetically-dominated plasmas   and apply it to explain gamma-ray flares from the  Crab Nebula \citep{2016arXiv160305731L}. The model relies on development of current-driven instabilities on macroscopic scales (not related to plasma skin depths), driven by  large-scales magnetic stresses  (of the type ``parallel currents attract''). Using analytical and numerical methods (fluid and particle-in-cell simulations), we study a number of model problems involving merger of both current-carrying and zero total current magnetic flux tubes  in relativistic magnetically-dominated plasma: (i) we extend Syrovatsky's classical model of  explosive $X$-point collapse to magnetically-dominated plasmas; (ii) we consider  instability of  two-dimensional force-free system of magnetic
islands/flux tubes (2D ``ABC'' structures); (iii)  we consider merger of two  zero total poloidal current  magnetic  flux tubes. In all cases  regimes of spontaneous and driven evolution are investigated.

\subsection{Large scale dynamics of PWNe - formation of current-carrying flux tubes }

The powerful Crab flares require that energy from a macroscopic scale is made available to the acceleration process.  As we will discuss in the following, current-driven MHD instabilities like the coalescence of parallel currents and the "X-point collapse" can be a viable way to achieve this.  One of the key  questions is then how a violently unstable and highly magnetized configuration is setup in the first place.  
To locate potential sites for the flaring region, we need to identify regions of high sigma and analyze the flow structure in these candidate flaring regions.  Here we use the result of simulations by \cite{2014MNRAS.438..278P}.
Although the scales required for the ``daily'' flare duration are not resolved by the simulation (its resolution is $>3$ light-days), it is instructive to correlate the high-sigma region (that forms as a consequence of flow-expansion) with the current distribution in the simulations.

Importantly, highly magnetized regions  in the bulk of the nebula  can be achieved via flow expansion at  intermediate latitude regions {\it starting with only mildly magnetized wind}. In  Fig. \ref{point} we show the magnetization in the $xy$-plane from simulations of \citep{2014MNRAS.438..278P}.  One can clearly see that the magnetization rises well above the maximal injected value of $\sigma=1$.  The highest value of $\sigma\approx8$ in the snapshot is achieved at the point where radial expansion reverses and forms the plume-like polar flow.

\begin{figure}[htbp]
\begin{center}
\includegraphics[height=10cm]{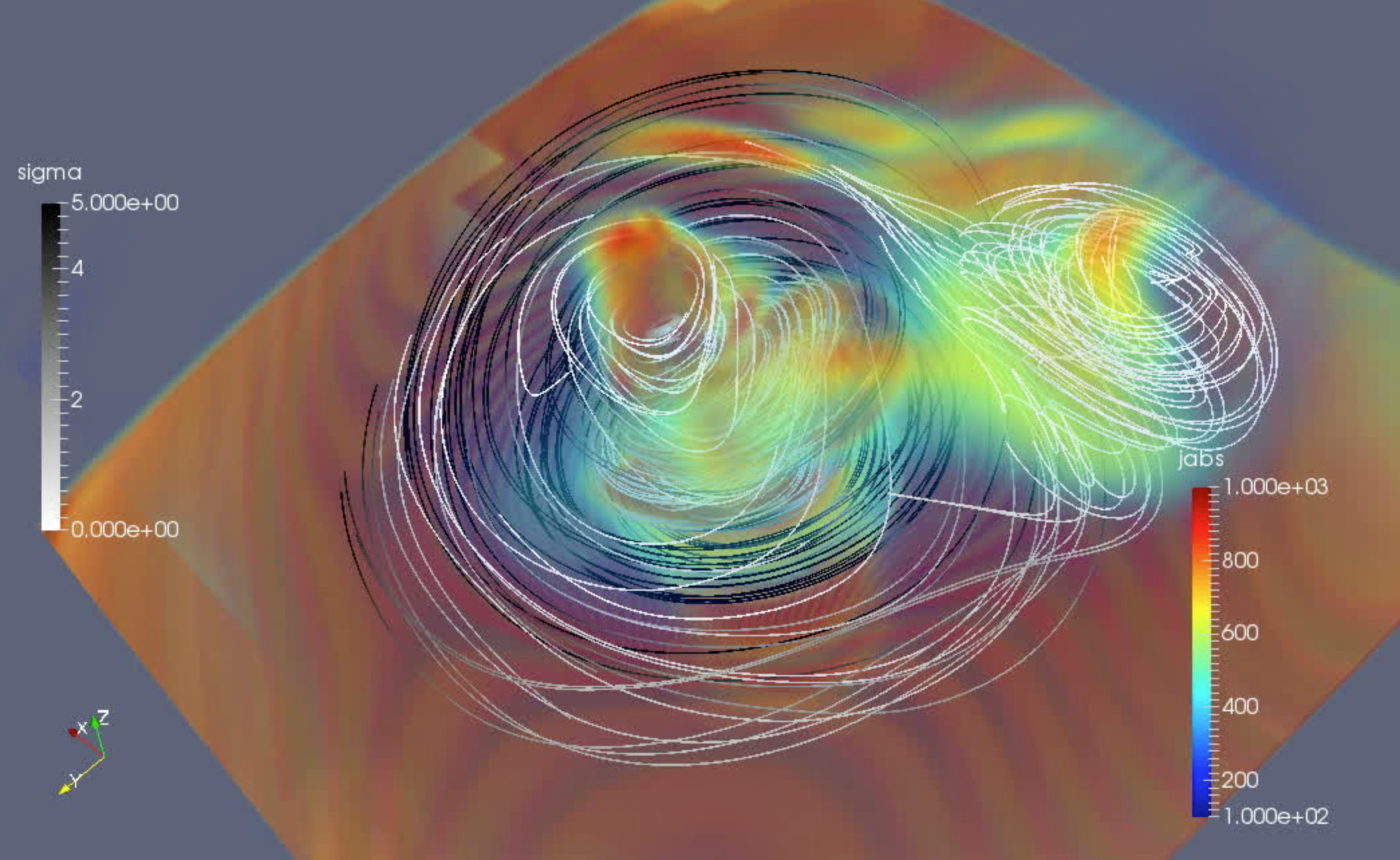}
\caption{3D volume rendering showing current filamentation of the polar beam just downstream of the termination shock.  The shock surface is indicated as the orange plane and we draw field-lines shaded from white $(\sigma=0)$ to black ($\sigma=5$).  
One clearly sees two current filaments producing structures similar to magnetic flux tubes.  As discussed in \cite{2014MNRAS.438..278P}, streamlines from intermediate latitudes reach the axis behind this inner violently unstable region and form a plume-like outflow of moderate velocity $v\approx0.7c$.  
}
\label{fig:filamentation}
\end{center}
\end{figure}

To better understand the geometry of the current and magnetic field, we display a representative volume rendering of the polar region in Fig. \ref{fig:filamentation}. 
In the rendering, one can see the violently unstable polar beam embedded into the more regular high-$\sigma$ region comprised of toroidal field lines.  
The plume forms downstream of this structure and is also strongly perturbed.  A part of the disrupted plume approaches the termination shock as a flux-tube.  
The presence of such a configuration where two flux tubes can come together very close to the high sigma region lets us speculate that Crab flares might originate when the right geometry (e.g. parallel flux tubes) coincides with high magnetization as present in the nebula even for moderate wind magnetization.  
For higher magnetizations of the polar beam, the mechanism described by \cite{2012MNRAS.427.1497L} could directly act also without first having to rely on enhancement of $\sigma$ via flow expansion.  Extrapolating from the moderate sigma simulations where the polar beam is highly unstable and forms a filamented current, this seems feasible at the very least.  

 As an initial pre-flare state of plasma we consider a 2D force-fee lattice of magnetic flux tubes 
 \ba &&
B_x = -\sin(2\pi \alpha y)B_0\,,
\nn &&
B_y = \sin(2\pi \alpha x) B_0 \,,
\nn &&
B_z =\left(\cos(2\pi \alpha x)+\cos(2\pi \alpha y)\right) B_0  \, ,
\label{Bb}
\ea
  This constitutes  a lattice of force-free magnetic islands separated by 
$90^o$ X-points in equilibrium. Islands have alternating out-of-the-plane poloidal  fields and  alternating toroidal fields. Each magnetic flux tube  carries a magnetic flux
$ \propto  B_0/\alpha^2$, energy per unit length $ \propto   B_0^2/\alpha^2$,  helicity per unit length $  \propto  B_0^2/(\alpha^3 )$ and axial current $  \propto  B_0/( \alpha)$. Helicity of both types of flux tubes is of the same sign.  Previously,  this configuration (called ABC) has been considered by \cite{1983ApJ...264..635P} in the context of Solar \Bfs

\begin{figure}[h!]
\centering
\includegraphics[width=.8\textwidth]{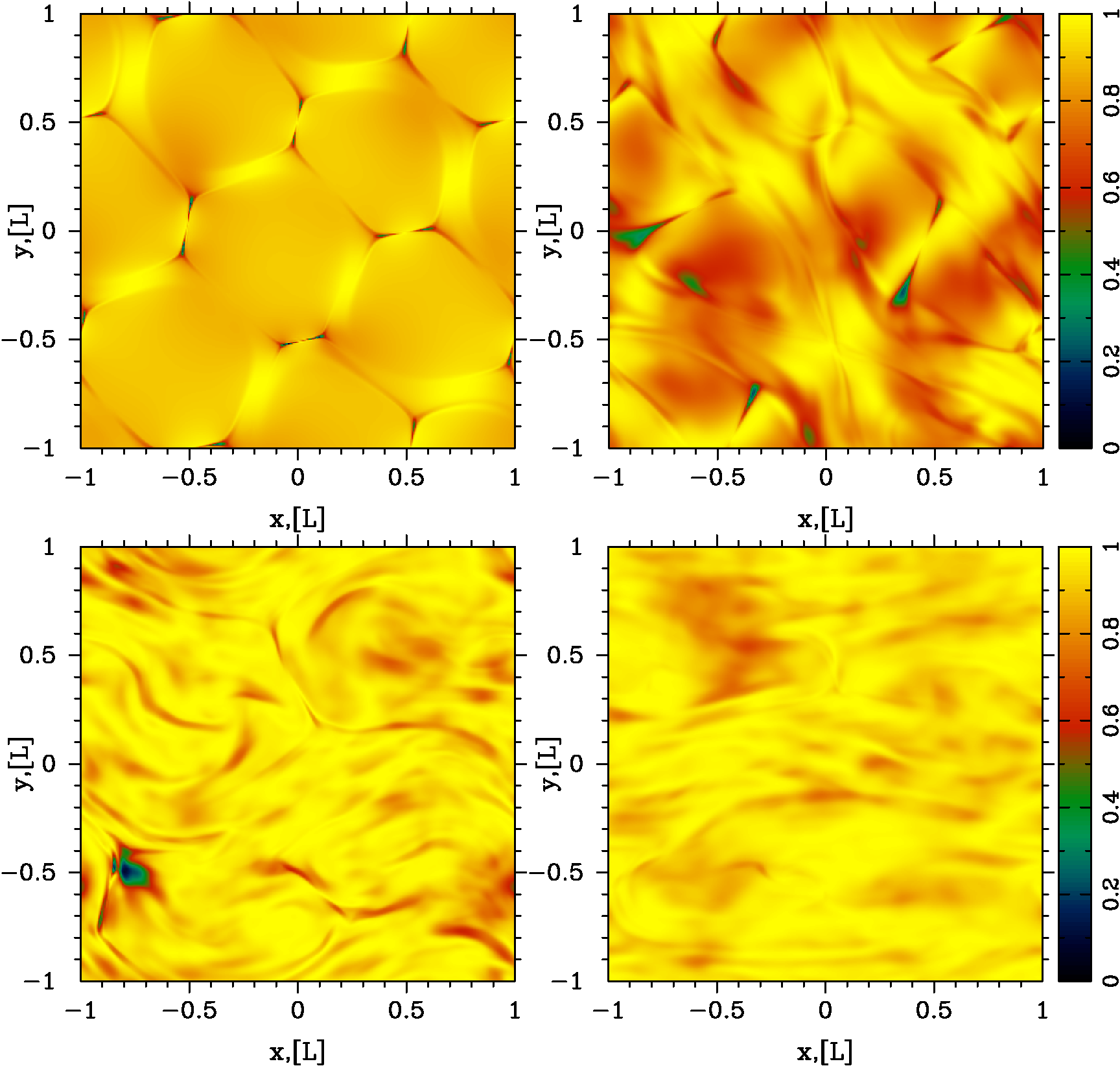}
\caption{X-point collapse and island merging for a set of
 unstressed magnetic islands in force-free simulations. 
We plot $1-E^2/B^2$ at times $t=8.0, 10.0, 15.0$ and 20.  
Compare with results of PIC simulations, Fig.  \protect\ref{fig:abcaccfluid}.
}
\label{ff-bme}
\end{figure}

The configuration (\ref{Bb})  is unstable, Fig. \ref{ff-bme}. The  instability of the 2D ABC configuration is of the kind ``parallel currents attract". In the initial configuration the attraction of parallel currents is balanced by the repulsion of anti-parallel ones.  Small amplitude fluctuations lead to fluctuating forces between the currents, that eventually lead to the disruption of the system. To identify the dominant instability mode let us consider a simplified model problem replacing each island by a solid tube carrying a given current. 
Such incompressible-type approximation is expected to be valid at early times, when the resulting motions are slow and the amount of the dissipated  magnetic energy is small.

 \begin{figure}[h!]
 \centering
\includegraphics[width=.33\textwidth]{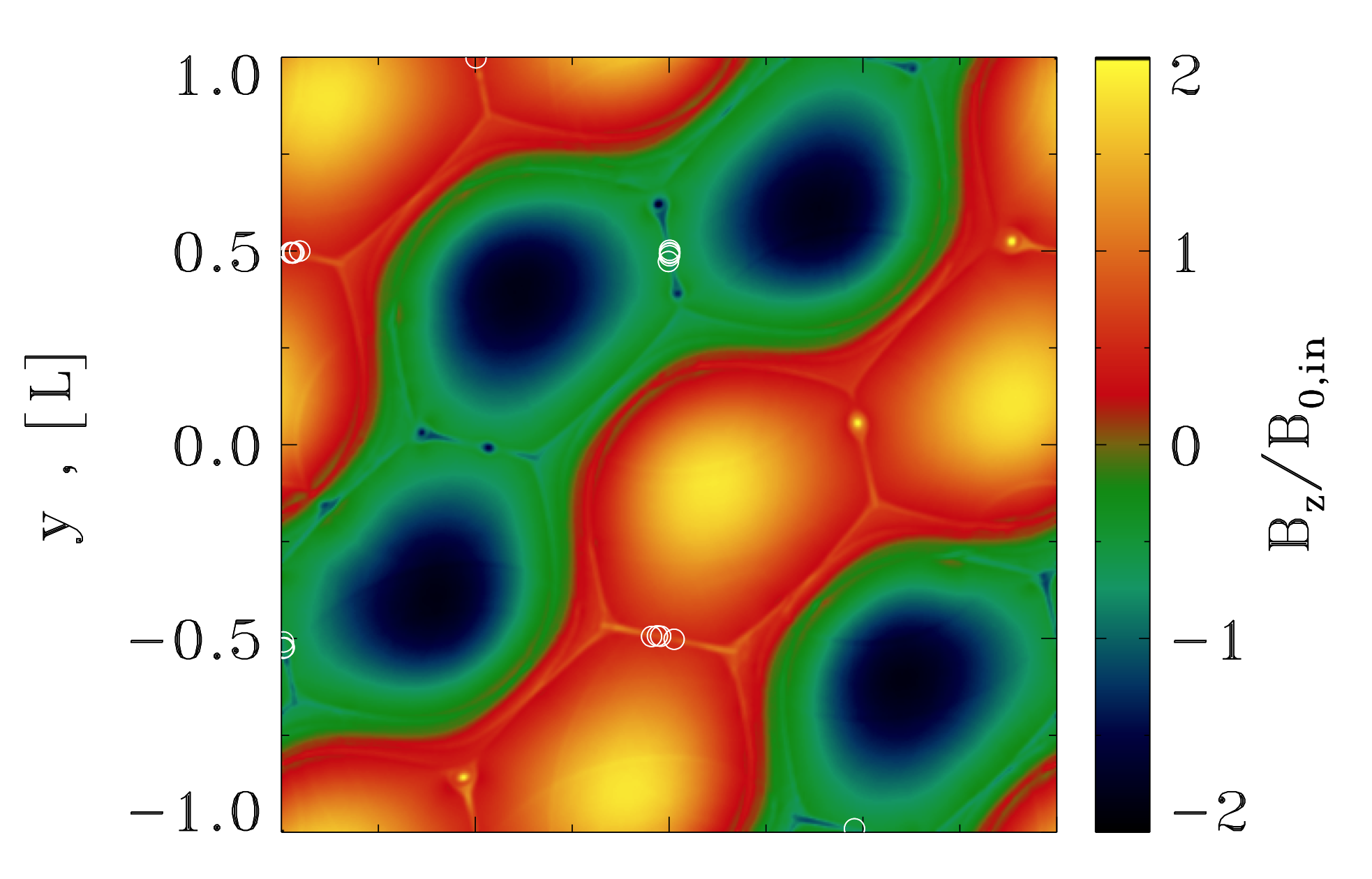} 
\includegraphics[width=.33\textwidth]{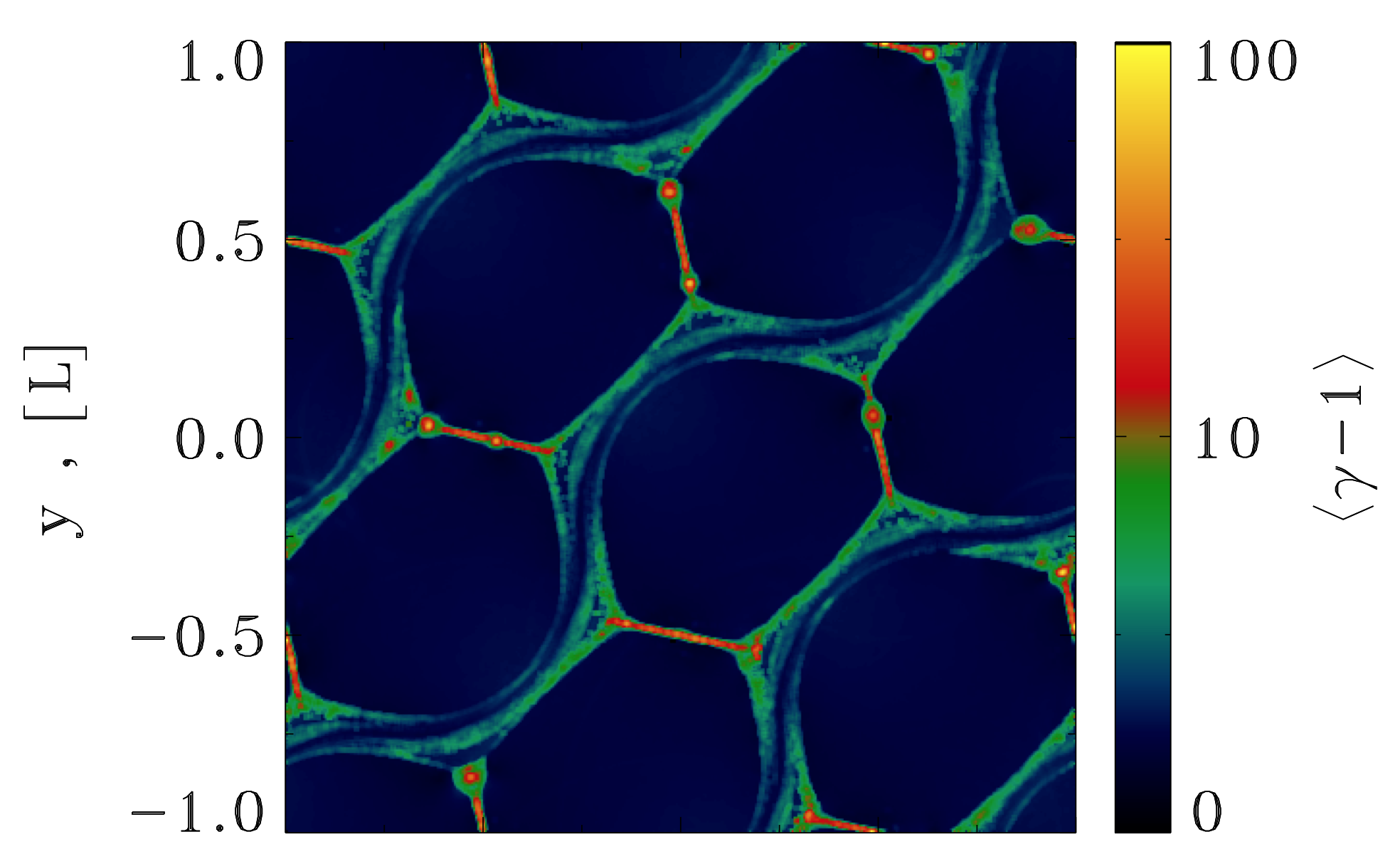} 
\includegraphics[width=.32\textwidth]{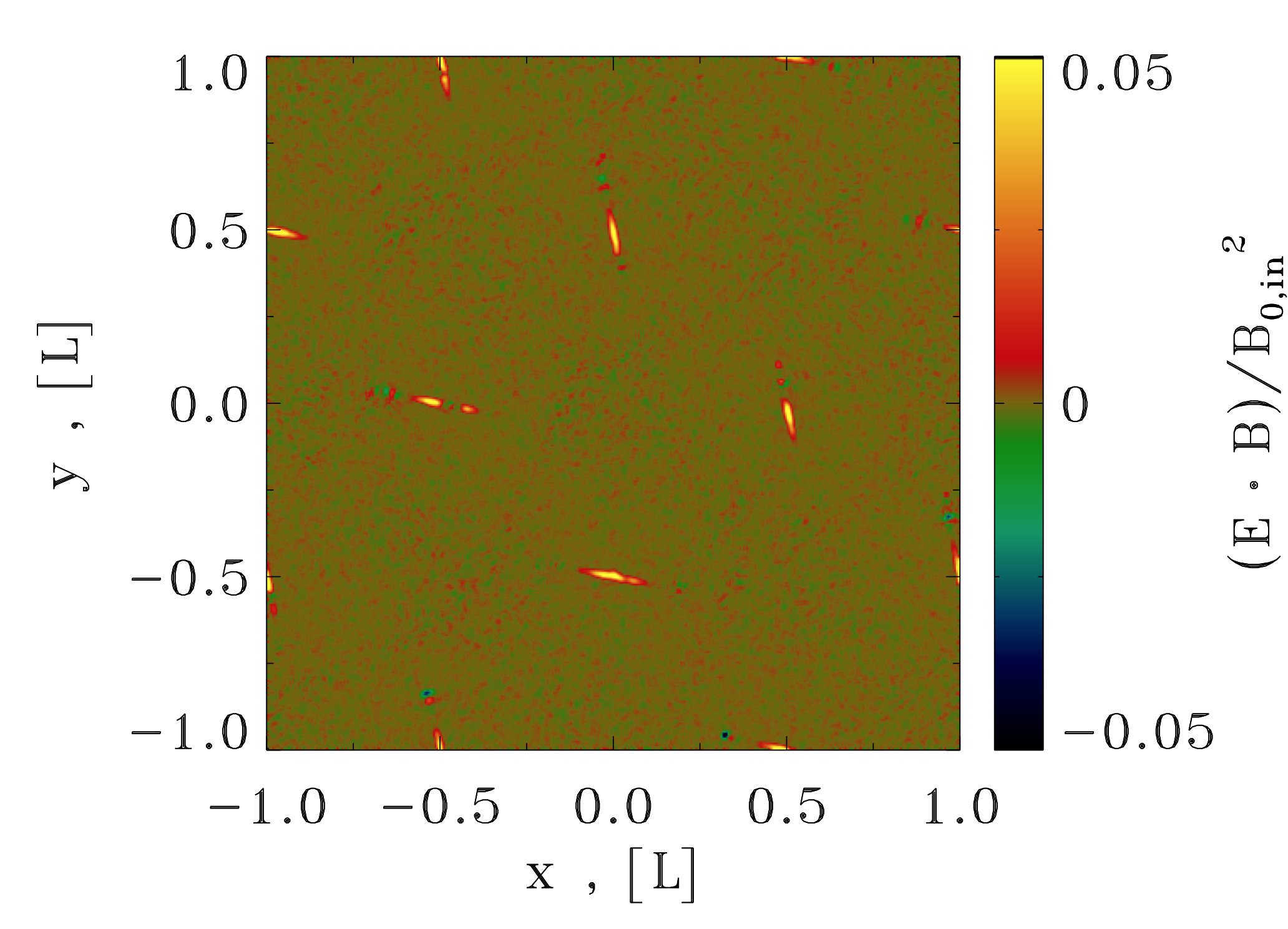} 
\caption{Physics of particle injection into the acceleration process, from  a 2D PIC simulation of ABC instability with $kT/mc^2=10^{-4}$, $\sigma=42$, performed within a square domain of size $2L\times 2L$. We plot the 2D ABC structure at $ct/L=6.65$. Left panel: 2D plot of the out-of-plane field $B_z$, in units of $B_{0,\rm in}$. Among the particles that exceed the threshold $\gamma_{\rm 0}=30$ within the interval $6.5\leq ct_0/L\leq6.8$ we select the 20 particles that at the final time reach the highest energies, and
 with open white circles we plot their locations at the injection time $t_0$. Center panel: 2D plot of the mean kinetic energy per particle $\langle\gamma-1\rangle$. Right panel: 2D plot of ${\bf E}\cdot{\bf B}/B_{0,\rm in}^2$, showing in red and yellow the regions of charge starvation. Comparison of the top panel with the bottom panel shows that particle injection is localized in the charge-starved regions.}
\label{fig:abcaccfluid} 
\end{figure}

We identify two stages of particle acceleration: (i)  fast explosive prompt X-point collapse and (ii)  ensuing  island merger. 
The fastest acceleration occurs during the initial catastrophic  X-point collapse, with the reconnection \Ef\ of the order
of the \Bf.   During the X-point collapse particles are accelerated by charge-starved \Efs, which can reach (and even exceed) values of  the local \Bf. The explosive stage of reconnection 
 produces non-thermal power-law tails with slopes
that depend on the average magnetization $\sigma$. For plasma magnetization $\sigma \leq 10^2$ the  spectrum  power law index is  $p< 2$; in this case   the maximal energy depends linearly on the size of the reconnecting islands. For  higher magnetization, $\sigma \geq 10^2$, the spectra are soft, $p< 2$, yet the maximal energy $\gamma_{max}$ can still exceed the average magnetic energy per particle,   $ \sim \sigma$,  by  orders of magnitude (if $p$ is not too close to unity).  
 The X-point collapse stage is followed by  magnetic island merger  that dissipates a large
fraction of the initial magnetic energy in a regime of forced magnetic
reconnection, further accelerating the particles, but proceeds at a slower reconnection rate.

Our model of Crab flares has a number of key features, that are both required by observations and/or have not been previously explored. 
\begin{description} 

\item {\it  Acceleration mechanism.} We argued that the particles producing Crab flares are accelerated in explosive magnetic reconnection events. This is, arguably, the first solidly established case in high energy astrophysics of direct acceleration in reconnection events  (as opposed to shock acceleration). In addition, since in our mode  the maximal energy that  particles can achieve grows with the size of the acceleration region, it is possible that smaller {\it reconnection events  are responsible for the acceleration of the majority of high-energy emitting particles in the Crab Nebula};  shock acceleration does work - producing the Crab inner knot \citep{2016MNRAS.456..286L} - but it may be subdominant for the acceleration of high energy particles.

\item {\it  Location of  flares.}  The flare-producing region is located 
at polar intermediate latitudes, between 10 and $\sim 45$ degrees, where the wind magnetization is expected to be high (the lower limit on the flare latitude comes from available required potential, while upper limit comes from modeling of the Crab inner knot \cite{2016MNRAS.456..286L}). The sectors of the wind that eventually become the  acceleration sites for flare particle have mild magnetization, 
$\sigma_w \sim 10-100$. Magnetization first increases at the oblique termination shock and later  in the bulk,  during the deceleration of the mildly relativistic post-shock flow.  As the flow decelerates to sub-relativistic velocities, large scale kink instabilities lead to formation of current-carrying flux tubes, Fig. \ref{fig:filamentation}.   
  
\item{\it  Size of the accelerating region.} In our model the acceleration occurs on {\it macroscopic scales, not related to the plasma microscopic scales, like the skin depth}.   
\citep[Previous models of reconnection in Crab flares, \eg][were based on the development of the tearing mode and achieved acceleration on scale related to the skin depth - there is not enough potential  on scales of few skin depths to account for Crab flares.]{2012ApJ...746..148C,2012ApJ...754L..33C} 

 \item   {\it Relativistic beaming motion of the flare producing region.} The peak frequency of flares, the energy  and the energetics of flares   all require mildly relativistic ``bulk'' motion of the flare producing particles, with $\Gamma \sim $ few. This is achieved via  ``kinetic beaming'', and not through a genuine, fluid-like bulk motion of the lower energy component.)

\end{description} 

 This work was supported by   NSF  grant AST-1306672 and DoE grant DE-SC0016369.

\bibliographystyle{apj}

\bibliography{/Users/maxim/Home/Research/BibTex}

\begin{thebibliography}{32}
\expandafter\ifx\csname natexlab\endcsname\relax\def\natexlab#1{#1}\fi

\bibitem[{{Abdo} {et~al.}(2011){Abdo}, {Ackermann}, {Ajello}, {Allafort},
  {Baldini}, {Ballet}, {Barbiellini}, \& {Bastieri}}]{2011Sci...331..739A}
{Abdo}, A.~A., {Ackermann}, M., {Ajello}, M., {et~al.} 2011, Science, 331, 739

\bibitem[{{Amano} \& {Kirk}(2013)}]{2013ApJ...770...18A}
{Amano}, T., \& {Kirk}, J.~G. 2013, \apj, 770, 18

\bibitem[{{Buehler} {et~al.}(2012){Buehler}, {Scargle}, {Blandford}, {Baldini},
  {Baring}, {Belfiore}, {Charles}, {Chiang}, {D'Ammando}, {Dermer}, {Funk},
  {Grove}, {Harding}, {Hays}, {Kerr}, {Massaro}, {Mazziotta}, {Romani}, {Saz
  Parkinson}, {Tennant}, \& {Weisskopf}}]{2012ApJ...749...26B}
{Buehler}, R., {Scargle}, J.~D., {Blandford}, R.~D., {et~al.} 2012, \apj, 749,
  26

\bibitem[{{B{\"u}hler} \& {Blandford}(2014)}]{2014RPPh...77f6901B}
{B{\"u}hler}, R., \& {Blandford}, R. 2014, Reports on Progress in Physics, 77,
  066901

\bibitem[{{Camus} {et~al.}(2009){Camus}, {Komissarov}, {Bucciantini}, \&
  {Hughes}}]{2009MNRAS.400.1241C}
{Camus}, N.~F., {Komissarov}, S.~S., {Bucciantini}, N., \& {Hughes}, P.~A.
  2009, MNRAS, 400, 1241

\bibitem[{{Cerutti} {et~al.}(2012{\natexlab{a}}){Cerutti}, {Uzdensky}, \&
  {Begelman}}]{2012ApJ...746..148C}
{Cerutti}, B., {Uzdensky}, D.~A., \& {Begelman}, M.~C. 2012{\natexlab{a}},
  \apj, 746, 148

\bibitem[{{Cerutti} {et~al.}(2012{\natexlab{b}}){Cerutti}, {Werner},
  {Uzdensky}, \& {Begelman}}]{2012ApJ...754L..33C}
{Cerutti}, B., {Werner}, G.~R., {Uzdensky}, D.~A., \& {Begelman}, M.~C.
  2012{\natexlab{b}}, \apjl, 754, L33

\bibitem[{{Clausen-Brown} \& {Lyutikov}(2012)}]{2012MNRAS.426.1374C}
{Clausen-Brown}, E., \& {Lyutikov}, M. 2012, MNRAS, 426, 1374

\bibitem[{{de Jager} {et~al.}(1996){de Jager}, {Harding}, {Michelson}, {Nel},
  {Nolan}, {Sreekumar}, \& {Thompson}}]{1996ApJ...457..253D}
{de Jager}, O.~C., {Harding}, A.~K., {Michelson}, P.~F., {et~al.} 1996, \apj,
  457, 253

\bibitem[{{Hester} {et~al.}(1995){Hester}, {Scowen}, {Sankrit}, {Burrows},
  {Gallagher}, {Holtzman}, {Watson}, {Trauger}, {Ballester}, {Casertano},
  {Clarke}, {Crisp}, {Evans}, {Griffiths}, {Hoessel}, {Krist}, {Lynds},
  {Mould}, {O'Neil}, {Stapelfeldt}, \& {Westphal}}]{1995ApJ...448..240H}
{Hester}, J.~J., {Scowen}, P.~A., {Sankrit}, R., {et~al.} 1995, \apj, 448, 240

\bibitem[{{Hester} {et~al.}(2002){Hester}, {Mori}, {Burrows}, {Gallagher},
  {Graham}, {Halverson}, {Kader}, {Michel}, \& {Scowen}}]{2002ApJ...577L..49H}
{Hester}, J.~J., {Mori}, K., {Burrows}, D., {et~al.} 2002, \apjl, 577, L49

\bibitem[{{Kennel} \& {Coroniti}(1984)}]{kc84}
{Kennel}, C.~F., \& {Coroniti}, F.~V. 1984, \apj, 283, 694

\bibitem[{{Komissarov}(2013)}]{2013MNRAS.428.2459K}
{Komissarov}, S.~S. 2013, MNRAS, 428, 2459

\bibitem[{{Komissarov} \& {Lyubarsky}(2004)}]{KomissarovLyubarsky}
{Komissarov}, S.~S., \& {Lyubarsky}, Y.~E. 2004, MNRAS, 349, 779

\bibitem[{{Lang} \& {Gingerich}(1979)}]{1979sbaa.book.....L}
{Lang}, K.~R., \& {Gingerich}, O. 1979, {A source book in astronomy and
  astrophysics, 1900-1975}

\bibitem[{{Lyubarsky}(2002)}]{2002MNRAS.329L..34L}
{Lyubarsky}, Y.~E. 2002, MNRAS, 329, L34

\bibitem[{{Lyubarsky}(2003)}]{2003MNRAS.345..153L}
---. 2003, MNRAS, 345, 153

\bibitem[{{Lyubarsky}(2012)}]{2012MNRAS.427.1497L}
---. 2012, MNRAS, 427, 1497

\bibitem[{{Lyutikov}(2010)}]{2010MNRAS.405.1809L}
{Lyutikov}, M. 2010, MNRAS, 405, 1809

\bibitem[{{Lyutikov} {et~al.}(2016{\natexlab{a}}){Lyutikov}, {Komissarov}, \&
  {Porth}}]{2016MNRAS.456..286L}
{Lyutikov}, M., {Komissarov}, S.~S., \& {Porth}, O. 2016{\natexlab{a}}, \mnras,
  456, 286

\bibitem[{{Lyutikov} \& {Lazarian}(2013)}]{2013SSRv..178..459L}
{Lyutikov}, M., \& {Lazarian}, A. 2013, \ssr, 178, 459

\bibitem[{{Lyutikov} {et~al.}(2016{\natexlab{b}}){Lyutikov}, {Sironi},
  {Komissarov}, \& {Porth}}]{2016arXiv160305731L}
{Lyutikov}, M., {Sironi}, L., {Komissarov}, S., \& {Porth}, O.
  2016{\natexlab{b}}, ArXiv e-prints

\bibitem[{{Minkowski}(1942)}]{1942ApJ....96..199M}
{Minkowski}, R. 1942, \apj, 96, 199

\bibitem[{{Parker}(1983)}]{1983ApJ...264..635P}
{Parker}, E.~N. 1983, \apj, 264, 635

\bibitem[{{P{\'e}tri} \& {Lyubarsky}(2007)}]{2007A&A...473..683P}
{P{\'e}tri}, J., \& {Lyubarsky}, Y. 2007, \aap, 473, 683

\bibitem[{{Porth} {et~al.}(2014){Porth}, {Komissarov}, \&
  {Keppens}}]{2014MNRAS.438..278P}
{Porth}, O., {Komissarov}, S.~S., \& {Keppens}, R. 2014, MNRAS, 438, 278

\bibitem[{{Rees} \& {Gunn}(1974)}]{reesgunn}
{Rees}, M.~J., \& {Gunn}, J.~E. 1974, MNRAS, 167, 1

\bibitem[{{Rudy} {et~al.}(2015){Rudy}, {Horns}, {DeLuca}, {Kolodziejczak},
  {Tennant}, {Yuan}, {Buehler}, {Arons}, {Blandford}, {Caraveo}, {Costa},
  {Funk}, {Hays}, {Lobanov}, {Max}, {Mayer}, {Mignani}, {O'Dell}, {Romani},
  {Tavani}, \& {Weisskopf}}]{2015ApJ...811...24R}
{Rudy}, A., {Horns}, D., {DeLuca}, A., {et~al.} 2015, \apj, 811, 24

\bibitem[{{Sironi} \& {Spitkovsky}(2011)}]{2011ApJ...741...39S}
{Sironi}, L., \& {Spitkovsky}, A. 2011, \apj, 741, 39

\bibitem[{{Tavani} {et~al.}(2011){Tavani}, {Bulgarelli}, {Vittorini},
  {Pellizzoni}, {Striani}, {Caraveo}, {Weisskopf}, \&
  {Tennant}}]{2011Sci...331..736T}
{Tavani}, M., {Bulgarelli}, A., {Vittorini}, V., {et~al.} 2011, Science, 331,
  736

\bibitem[{{Weisskopf} {et~al.}(2000){Weisskopf}, {Hester}, {Tennant}, {Elsner},
  {Schulz}, {Marshall}, {Karovska}, {Nichols}, {Swartz}, {Kolodziejczak}, \&
  {O'Dell}}]{2000ApJ...536L..81W}
{Weisskopf}, M.~C., {Hester}, J.~J., {Tennant}, A.~F., {et~al.} 2000, \apjl,
  536, L81

\bibitem[{{Weisskopf} {et~al.}(2013){Weisskopf}, {Tennant}, {Arons},
  {Blandford}, {Buehler}, {Caraveo}, {Cheung}, {Costa}, {de Luca}, {Ferrigno},
  {Fu}, {Funk}, {Habermehl}, {Horns}, {Linford}, {Lobanov}, {Max}, {Mignani},
  {O'Dell}, {Romani}, {Striani}, {Tavani}, {Taylor}, {Uchiyama}, \&
  {Yuan}}]{2013ApJ...765...56W}
{Weisskopf}, M.~C., {Tennant}, A.~F., {Arons}, J., {et~al.} 2013, \apj, 765, 56

\end{thebibliography}

\end{document}